\documentclass[superscriptaddress,twocolumn,floatfix,showkeys,preprintnumbers,showpacs]{revtex4}
\usepackage{amscd}
\usepackage{epsfig}
\usepackage{amsmath}
\usepackage{graphicx}
\usepackage{amsfonts}
\usepackage{amssymb}

\newtheorem{lemma}{Lemma}

\newtheorem{corollary}{Corollary}

\begin{document}

\begin{verbatim}
\title{Entropy, dimension, and state mixing in a class of time-delayed
dynamical systems}
\author{D. J. Albers}
\email{albers@cse.ucdavis.edu}
\affiliation{Max Plank Institute for Mathematics in the Sciences,
Leipzig 04103, Germany}

\author{Fatihcan M. Atay}
\email{atay@member.ams.org}
\affiliation{Max Plank Institute for Mathematics in the Sciences,
Leipzig 04103, Germany}

\date{\today}
\keywords{Chaos, high dimensions, structural stability,
Lyapunov exponents, delay, entropy,
Kaplan-Yorke dimension}
\pacs{05.45.-a, 89.75.-k, 05.45.Tp, 02.30.Ks, 05.45.Jn, 05.45.Pq}

\begin{abstract}%
  Time-delay systems are, in many ways, a natural set of dynamical
  systems for natural scientists to study because they form an
  interface between abstract mathematics and data. However, they are
  complicated because past states must be sensibly incorporated into
  the dynamical system. The primary goal of this paper is to begin to
  isolate and understand the effects of adding time-delay coordinates
  to a dynamical system. The key results include (i) an analytical
  understanding regarding extreme points of a time-delay dynamical
  system framework including an invariance of entropy and the variance
  of the Kaplan-Yorke formula with simple time re-scalings; (ii)
  computational results from a time-delay mapping that forms a path
  between dynamical systems dependent upon the most distant and the
  most recent past; (iii) the observation that non-trivial mixing of
  past states can lead to high-dimensional, high-entropy dynamics that
  are not easily reduced to low-dimensional dynamical systems; (iv)
  the observed phase transition (bifurcation) between low-dimensional,
  reducible dynamics and high or infinite-dimensional dynamics; and
  (v) a convergent scaling of the distribution of Lyapunov exponents,
  suggesting that the infinite limit of delay coordinates in systems
  such are the ones we study will result in a continuous or (dense) point
  spectrum.%
\end{abstract}%

\maketitle
\end{verbatim}

\section{Introduction}

Experimental, scientific data for which time is an independent
parameter is collected in the form of a scalar or vector time-series.
The vector time-series rarely measures all of the independent
coordinates required for a full specification of the system; the
scalar time-series data never will. Nevertheless, that even a scalar
time-series can be used to represent and reconstruct the original
dynamical or natural system from which the data originated was a
problem addressed well by Takens \cite{takensstit}, Packard et
al.~\cite{geo_of_time_series}, and Sauer et al.~\cite{embedology}.
That there exist mappings that can reconstruct the dynamical system
from observed time-series has also been shown (e.g., Hornik et
al.~\cite{hor2}), even if the actual reconstruction has proven
difficult \cite{sprott_book,kantz_book}.  Nevertheless, it is usually
time-delay dynamical systems that are of prime interest for practical
analysis of natural systems because they are often the dynamical
systems closest to real data.  In this paper, we study discrete-time
dynamical systems wth time delays. There are, of course, many
formulations of time-delay dynamical systems; we wish to target and
isolate the effects associated with adding time-delay coordinates
using the simplest possible construction (for an alternative, see
\cite{manffra_kantz}, \cite{doyne_dds}, or \cite{clint_dde}). To
achieve this end, we have structured this paper so as to study various
extremes that are complimented with results for intermediate cases. In
particular, we consider the dynamics of an iterated map and its mixing
with a single delay from the distant past. Moreover, to isolate and
demonstrate the diversity among the different mappings, we consider
two maps whose parameter spaces are diametrically opposed --- the
logistic map, which has dense stable periodic orbits for positive
parameter values for which it remains bounded; and the tent map, which
has a unique Sinai-Ruelle-Bowen (SRB) measure \cite{youngSRB} over a
large portion of its parameter space.  It is worth noting that despite
this difference, these maps are conjugate to one another at least one
parameter value.  A fundamental computational analysis of delay dynamical
systems as they are used to approximate delay differential equations,
and the characteristics of the diagnostics we will also study is
presented in Ref. \cite{doyne_dds}, which provides the best
computational background for the study we will present in this paper.

We begin introducing time-delay systems in section \ref{sec:framework}
and the associated diagnostics in section \ref{sec:diagnostics}.  With
this groundwork laid, we begin the analysis in section
\ref{sec:analytics} with an analytical study of both the dynamics and
the diagnostics of two extremes --- (scalar) mappings dependent only
on the most recent time-step and mappings dependent only on a single
time-step from the distant past. Said differently, we study the
dynamics and isolate the effects on various standard diagnostics of a
simple time-rescaling where there is no mixing of states at different
times.  While we will claim some circumstances where the delay
dynamical systems we study approximate infinite-dimensional,
continuous spectrum systems, we are also interested in isolating the
effects of rescaling time and adding delay coordinates.  In the
circumstance when time is rescaled, we show that the metric entropy is
invariant to the time-rescaling, the largest Lyapunov exponent follows
a simple rescaling that is a function of the time-delay, and the
Kaplan-Yorke dimension formula can produce deceiving results. (It is
known that as the delay is increased, Kaplan-Yorke dimension increases
linearly; we will provide insight into why this is so.)  The section
\ref{sec:numerics} intermediate cases follow via a computational study
that forms a bridge between the normal and the time-rescaled maps. As
past states are mixed, for similar reasons that allow for the
time-series embedding theorems to function, the dynamics become much
more complicated and are not easily reduced to low-dimensional
dynamical systems. Moreover, as the states are mixed, the results can
depend on the parity of the number of delays and often depend
profoundly on the chosen map.  Aside from studying the effects of
adding delay coordinates, it will also prove important to study the
variation that exists over different explicit mappings.  To isolate
the effects of adding delays from the effects dependent on a
particular mapping, we consider, as previously mentioned, two
practical extrema among mappings --- the tent and logistic maps. These
maps represent ``functional extrema'' in the sense that the tent map
has a unique SRB measure for a large, hole-free, open set of parameter
values; whereas the logistic map has stable, hyperbolic, periodic
dynamics for a dense set of parameter values \cite{graczyk_paper}.
Thus, the tent map is extremely dynamically stable in the sense that
chaotic dynamics is maintained when parameters are changed. This is in
contrast to the logistic map which, upon parameter variation, bears
witness to catastrophic changes in dynamical behavior realized via the
dense stable periodic orbit structure in parameter space.
Nevertheless, we will observe that adding time-delays decreases, in a
broad sense, the existence of periodic windows even for maps that have
dense stable periodic windows in their parameter space. Moreover,
high-entropy, high-dimensional geometric structure is observed for
non-trivial mixing of previous states.

\section{Framework}
\label{sec:framework}

We address issues related to dynamical systems where the present
(time-delay-vector) state
\begin{equation*}
\mathbf{x}_{t}=(x_{t},x_{t-1},x_{t-2},\cdots,x_{t-\tau})
\end{equation*}
is dependent upon past states with mappings of the form:
\begin{align*}
\mathbf{x}_{t+1} &  =\mathbf{F}(\mathbf{x_{t}})\\
&  =(F(x_{t},\dots,x_{t-\tau}),x_{t},x_{t-1},\cdots,x_{t-\tau+1}%
)
\end{align*}
where $\mathbf{F},F\in C^{r}$ ($r>0$), $\tau\in\mathbb{N}$, $\mathbf{x_{t}}%
\in\mathbb{R}^{\tau+1}$, and $x_{t}\in\mathbb{R}$ is always bounded. There
exist an infinite number of ways to combine current and previous states, 
for instance by a simple summation of previous states represented by:
\begin{equation}
x_{t+1}=F(x_{t},\dots,x_{t-\tau})=\sum_{i=0}^{\tau}\beta_{i}f_{i}%
(x_{t-i})\label{equation:full-delay-sum}%
\end{equation}
where $\beta_{i}\in\mathbb{R}$ and $f\in C^{r}$ ($r>0$). One can further
restrict to the case where $f_{i}$ is identical for all $i$. One
nontrivial example worth mentioning where the $f_{i}$'s are \emph{not}
identical, but where $F$ remains a \emph{linear combination} of previous
states, is the standard \emph{delayed feedback} case which can be arrived at
by setting $f_{0}$ to a $C^{r}$ map, $(1-\beta_{\tau})f$, and $f_{\tau}%
=\beta_{\tau}x_{\tau}$ (see \cite{logistic_delayed_feedback} for more
information on this particular formulation). Note that all of the above
time-delay dynamical systems are $d=\tau+1$ dimensional.


In this paper we concentrate on the case
\begin{equation}
x_{t+1}=(1-\beta)f(x_{t})+\beta f(x_{t-\tau})\label{main}%
\end{equation}
for some given $f:\mathbb{R}\rightarrow\mathbb{R}$, where $\tau\in
\mathbb{Z}^{+}$ is the time delay and the scalar $\beta\in\lbrack0,1]$ is a
measure of the relative effect of the past on the evolution. 
With $\beta=0$ we have the
evolution generated by the simple iteration rule%
\begin{equation}
x_{t+1}=f(x_{t})\label{b0}%
\end{equation}
which corresponds to the standard map with no delays (ND), whereas when
$\beta=1$ we obtain what we will call a \emph{pure delay} (PD) system
\begin{equation}
x_{t+1}=f(x_{t-\tau}).\label{PD}%
\end{equation}
A primary question we will address is the nature of the change in dynamics of
(\ref{main}) between these two extreme (ND and PD) cases as $\beta$ is varied.
For $\beta\in(0,1),$ (\ref{main}) has a $(\tau+1)$-dimensional state space in
the coordinates $\mathbf{x}_{t}=(x_{t},x_{t-1},\dots,x_{t-\tau}),$ so it is
convenient to view also the extreme cases as $(\tau+1)$-dimensional. The
system (\ref{main}) thus provides a simple background for investigating the
effect of past information on the dynamics.

An important motivation for studying the system (\ref{main}) comes from
synchronization of networks. Indeed, the so-called coupled map lattice
\cite{Kaneko-book93} in the presence of transmission delays takes the form
\cite{Atay-PRL04}%
\begin{equation}
x_{t+1}^{i}=f(x_{t}^{i})+\frac{\beta}{k_{i}}\sum a_{ij}(f(x_{t-\tau}%
^{j})-f(x_{t}^{i})),\qquad i=1,\dots,N.\label{network}%
\end{equation}
Here $x_{t}^{i}$ is the state at time $t$ of the $i$th member (node)
of a network of $N$ coupled dynamical systems, each of which follows
the evolution rule (\ref{b0}) in isolation, but interacts with its
neighbors when coupled to the network. The scalar $\beta$ represents
the coupling strength. The scalars $a_{ij}$ are 1 whenever $i$ and $j$
are neighbors and zero otherwise, and $k_{i}=\sum_{j}a_{ij}$ is the
degree of node $i$, i.e., its number of neighbors. The nonnegative
integer $\tau$ represents the time delay in the information
transmission between the neighbors of the network. It has been shown
that the system (\ref{network}) can \emph{synchronize}, i.e.,
$|x_{t}^{i}-x_{t}^{j}|\rightarrow0$ as $t\rightarrow\infty$ for all
$i,j$, even in the presence of a positive transmission delay
\cite{Atay-PRL04}. Then (\ref{network}) asymptotically approaches a
synchronous solution where $x_{t}^{i}=x_{t}$ for all $i$. 
It is easy to see then that the synchronous solution $x_{t}$ satisfies
(\ref{main}). In other words, (\ref{main}) describes the dynamics of the
synchronous solutions of coupled map lattices in the presence of
transmission delays. It has been shown that the presence of delays
greatly enriches the synchronous dynamics, whereas in the undelayed case
the dynamics of the synchronized network and the isolated units are 
identical \cite{Atay-PRL04,Atay-Complexity04}. 
We investigate further aspects of this observation in the following
sections.


\section{Diagnostics}
\label{sec:diagnostics}

The primary diagnostic quantities we will use in this paper are the Lyapunov
characterisctic exponents (LCE) and quantities defined by the 
Lyapunov spectrum, such as the
metric entropy $h_{\mu}$ and the Kaplan-Yorke dimension, $D_{KY}$
\cite{pesinlebook,shimnag}. Recall that each Lyapunov exponent in the
Lyapunov spectrum is given by:
\begin{equation}
\chi_{j}=\lim_{t\rightarrow\infty}\frac{1}{t}\ln\Vert((D\mathbf{F}^{t})^{\top
}(D\mathbf{F}^{t}))^{1/2}\cdot v_{j}\Vert\label{equation:LCEs}%
\end{equation}
where $(D\mathbf{F}^{t})^{\top}$ is the transpose of the Jacobian 
$D\mathbf{F}^{t}$
and $v_{j}$ is a basis element of the tangent space (i.e., there are
$d$, $d$-dimensional, mutually orthogonal vectors, each of which
correspond to basis elements of the tangent space; for more
information, see \cite{pesinlebook,ruellehilbert,guck}). 
For convenience, we will assume that the Lyapunov
exponents are {monotonically} ordered by index according to
$\chi_{i}\geq\chi_{i+1}$. 
In this work, we utilize the standard algorithm for computing the LCEs numerically as is given
in Benettin et al.~\cite{benn2} or Shimada and Nagashima
\cite{shimnag}. 
Furthermore, the metric entropy is given by the sum of the positive
LCEs,
\begin{equation}
h_{\mu}=\sum_{\chi_{i}>0}\chi_{i}.
\label{entropy} 
\end{equation} 
Similarly, the Kaplan-Yorke
dimension of an attractor \cite{Kaplanyorke,kaplan_yorke_2} is given
by:
\begin{equation}
D_{KY}=j+\frac{\chi_{d}+\dots+\chi_{d-j}}{|\chi_{d-j-1}|} \label{equation:kyc}%
\end{equation}
where $j<d$ is the largest integer such that $\chi_{d}+\dots+\chi_{d-j}\geq0$.



\section{Effects of a pure delay-time rescaling}
\label{sec:analytics}

The ND and PD cases represent the extrema of Eq.
(\ref{equation:full-delay-sum}) relative to the $\beta$ parameters;
thus, an understanding of both the trivial and PD cases will form a
foundation for studying Eq. (\ref{equation:full-delay-sum})
and, in particular, Eq. (\ref{main}).
In this special case
the LCE scalings with $d$ can be handled analytically.
Thus, the following results apply for $f \in C^{r}$ ($r \geq1$),
assuming that $f$ supports a unique SRB measure \cite{youngSRB} or has
\emph{robust chaos} \cite{yorkerobustchaos,unimodalrobust}
(thus, these results are largely independent of a particular choice of
$f$).

\begin{lemma}
[Lyapunov spectrum for PD]\label{lemmma:lle_rescaling} The Lyapunov 
spectrum of (\ref{PD}) is  
\begin{equation}
\chi_{1}=\chi_{2}=\cdots=\chi_{\tau+1}=\frac{\mu}{\tau+1},
\label{LCE-PD}
\end{equation}
where $\mu$ is the Lyapunov exponent of $f$.
\end{lemma}

\emph{Proof.} Defining the vector $\mathbf{x}_{t}=(x_{t},x_{t-1},\dots,x_{t-\tau})\in\mathbb{R}^{\tau+1}$,
we write (\ref{PD}) in vector form 
\begin{equation}
\mathbf{x}_{t+1}=(f(x_{t-\tau}),x_{t},\dots,x_{t-\tau+1}).
\label{vectorPD}
\end{equation}
It follows that 
\begin{align*}
\mathbf{x}_{t+\tau+1} & = (x_{t+\tau+1},x_{t+\tau},\dots,x_{t+1})\\
                      & = (f(x_{t}),f(x_{t-1}),\dots,f(x_{t-\tau})).    
\end{align*}
Rescaling time as 
\begin{equation}
s=\frac{t}{\tau+1}
\label{timescale}
\end{equation}
gives
\begin{align*}
& (x_{(\tau+1)s+(\tau+1)},x_{(\tau+1)s+\tau},\dots,x_{(\tau+1)s+1}) \\ 
 & = (f(x_{(\tau+1)s}),f(x_{(\tau+1)s-1}),\dots,f(x_{(\tau+1)s-\tau}))  
\end{align*}
Finally letting $u_{s}^{i}=x_{(\tau+1)s-(i-1)}$ for $i=1,\dots,\tau+1$,
we obtain 
\[
(u_{s+1}^{1},u_{s+1}^{2},\dots,u_{s+1}^{\tau+1})=(f(u_{s}^{1}),f(u_{s}^{2}),\dots,f(u_{s}^{\tau+1})).
\]
The last equation describes $\tau+1$ decoupled scalar systems each
of which evolves by the identical rule of the form (\ref{b0}); so
it has $\tau+1$ identical Lyapunov exponents. In view of the applied
time scaling (\ref{timescale}), it follows that the Lyapunov exponents
of (\ref{vectorPD}) are given by (\ref{LCE-PD}).



From Lemma \ref{lemmma:lle_rescaling} and Eq.~(\ref{entropy}), the following corollary is immediate.

\begin{corollary}
[Metric entropy invariant to a PD]\label{corollary:e_invariant} The standard 
metric entropy $h_{\mu}$ for the pure delay system (\ref{PD}) is 
independent of $\tau$.
\end{corollary}

On the other hand, Lemma \ref{lemmma:lle_rescaling} and 
Eq.~(\ref{equation:kyc}) imply that the Kaplan-Yorke dimension 
is $D_{KY}=d=\tau + 1$, which yields the following. 

\begin{corollary}
[$D_{KY}$ is not invariant to a PD]\label{corollary:dky_not_invariant} The
Kaplan-Yorke dimension formula is not invariant to $\tau$ 
in the pure delay system (\ref{PD}).
\end{corollary}

Why is Corollary \ref{corollary:dky_not_invariant} important? The pure
delay system (\ref{PD}) is equivalent to the non-scaled system (\ref{b0}) in every
way \emph{but} the calculated dimension. Moreover, for the
PD system the ``dimension'' scales linearly with the delay.
As we will see, for 
(\ref{main}) also, $D_{KY}\approx d$ persists for
$\beta$ being a significant distance from one, and only decreases as $\beta$
approaches one-half. But, as we will see, as $\beta$ is decreased from one or
increased from zero, a significant change in the \emph{dynamics}, as
quantified by the invariant density and the structure of the LCEs, remains
undetected in the dimension calculations. In particular, we will see a
transition between the trivial high-dimensional dynamics of the PD that is
easily reducible, and an irreducible manifestation of high-dimensional
dynamics with no significant impact seen in $D_{KY}$ versus $\beta$. Thus, we
claim that $D_{KY}$, and several other dimension estimates, can yield
deceiving results for some time-delay dynamical systems because the $D_{KY}$
has an implicit coordinate dependence.

Summarizing, in dynamical systems with a PD that has no mixing of states for
different times, the largest LCE is decreased by the factor that time has been
rescaled, the metric entropy is invariant to the time-rescaling, and the
Kaplan-Yorke dimension is equal to the factor by which time has been rescaled
(i.e., $\tau+1$).

\section{Effects of state-mixing via added delay coordinates}
\label{sec:numerics}

With the endpoints ($\beta = 0$ and 1) fully understood we can now begin
to piece together the transitional region where states are mixed according to
Eq.~(\ref{main}).
At this time a full
analytical understanding of this system is unavailable. 
Thus, for what follows, we will
be restricted to a computational study. Moreover, as previously
mentioned, these results, unlike those of the above section,
\emph{will} depend on the particular mapping; hence, the reason for an
investigation using two common but dynamically distinct maps, the
logistic and tent maps \footnote{While the tent and logistic maps are
  conjugate for certain parameter values (logistic at $a=4$ (c.f. Eq.
  \ref{equation:logistic}) and the tent map at $b=2$ (c.f. Eq.
  \ref{equation:tent})), the logistic map displays non-robust chaos
  for all but a single parameter value (where it remains bounded)
  whereas the tent map displays robust chaos for a large portion of
  its parameter space.}.

\subsection{Tent map}

\begin{figure}[tb]
\begin{center}%
\begin{tabular}
[c]{c}%
{\parbox[b]{3.5293in}{\begin{center}
\includegraphics[scale=0.65]{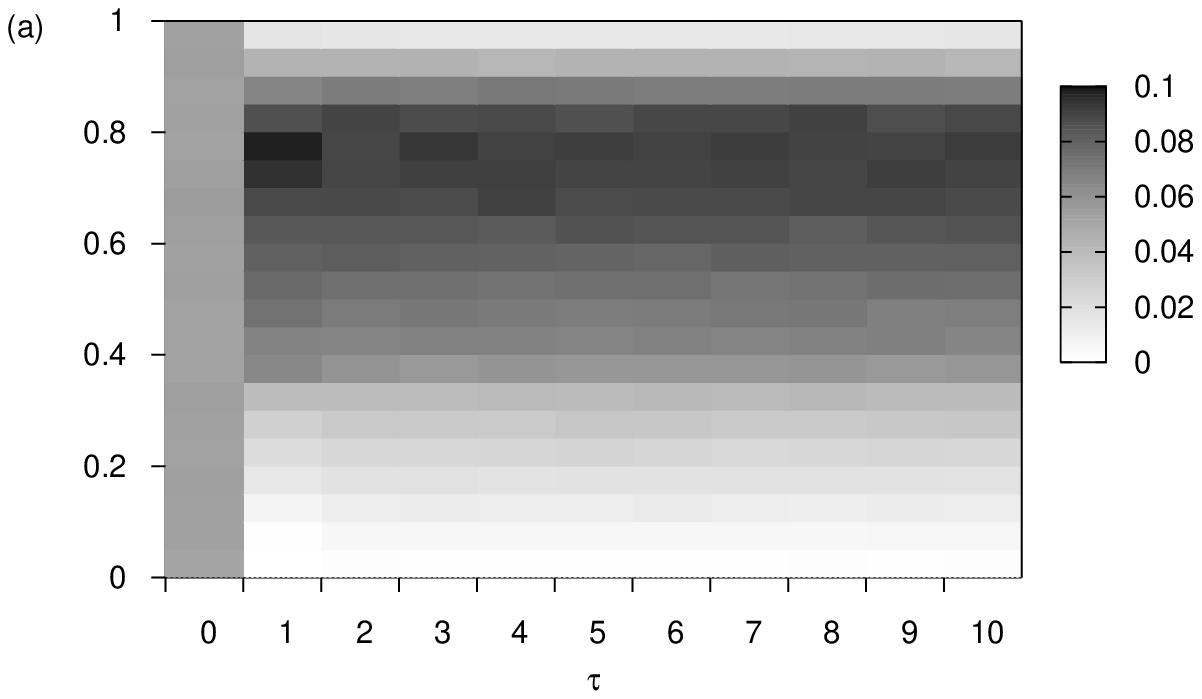}\end{center}}}\\
{\parbox[b]{3.5293in}{\begin{center}
\includegraphics[scale=0.65]{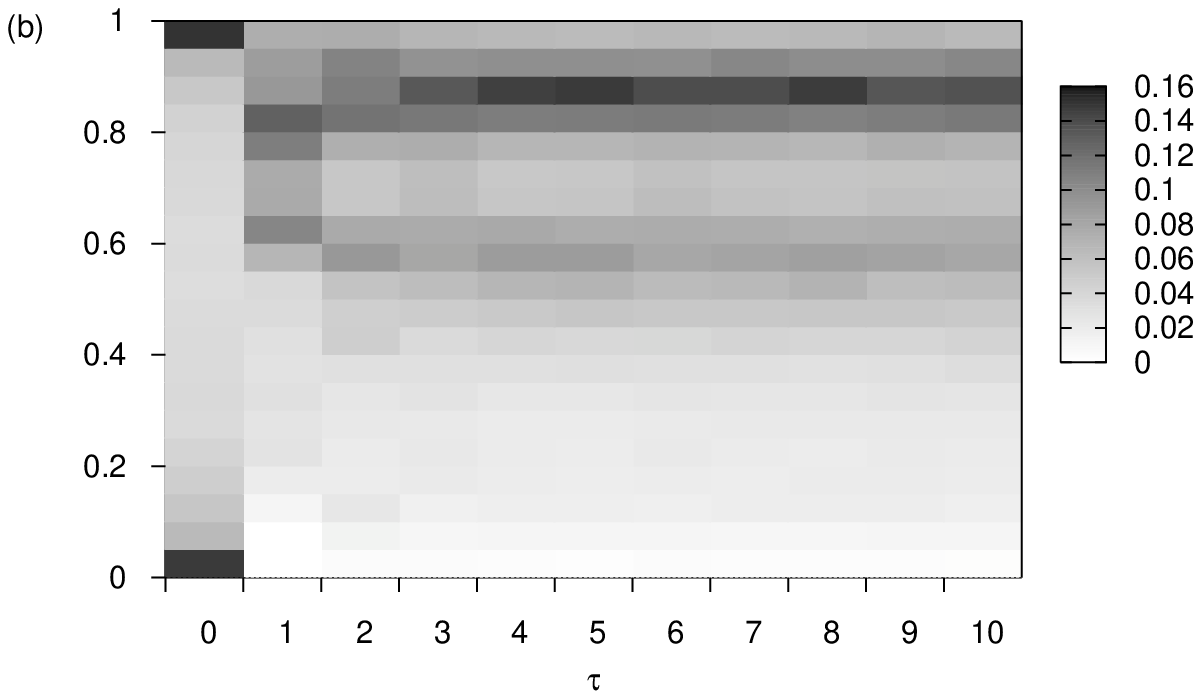}\end{center}}}%
\end{tabular}
\end{center}
\caption{The dependence of the natural density for (\ref{main}) on the delay
$\tau$. (a) The tent map, (b) the logistic map. $\beta=0.8$. }%
\label{fig:density2}%
\end{figure}

We will begin the computational analysis with the standard tent map
given by:
\begin{equation}
f(x)=\left\{
\begin{array}
[c]{ll}%
bx & \text{if $0<x\leq1/2$},\\
b-bx & \text{if $1/2\leq x\leq1$}.
\end{array}
\right.  \label{equation:tent}%
\end{equation}
at $b=2$. The first case we will consider is the tent map with $49$
delay coordinates ($d=50$) as this is a good intermediate value
between the low-$d$ and high-$d$ cases. Considering 
Fig.~\ref{fig:density2}, when $f$ is the standard tent map, there is little
difference in the qualitative structure of the map for $d>4$; 
for $d>4$, all dimensional dependence and parity
disappears (moreover, there do not exist periodic windows for $d>3$).
Nevertheless, when $d<5$, there is significant dynamical variation as
the parameters and the number of dimensions are changed. This dynamic
variation includes the existence of periodic windows in the $\beta
$-parameter space, dimension parity, and the lack of symmetry about
$\beta=1/2$. This dimensional cutoff is likely related to the rate of
decay of mutual information between $x_{t}$ and $x_{t-\tau}$; however,
a precise understanding of this ``functional'' bifurcation is yet to
be understood.

\begin{figure}[ptb]
\begin{center}
\epsfig{file=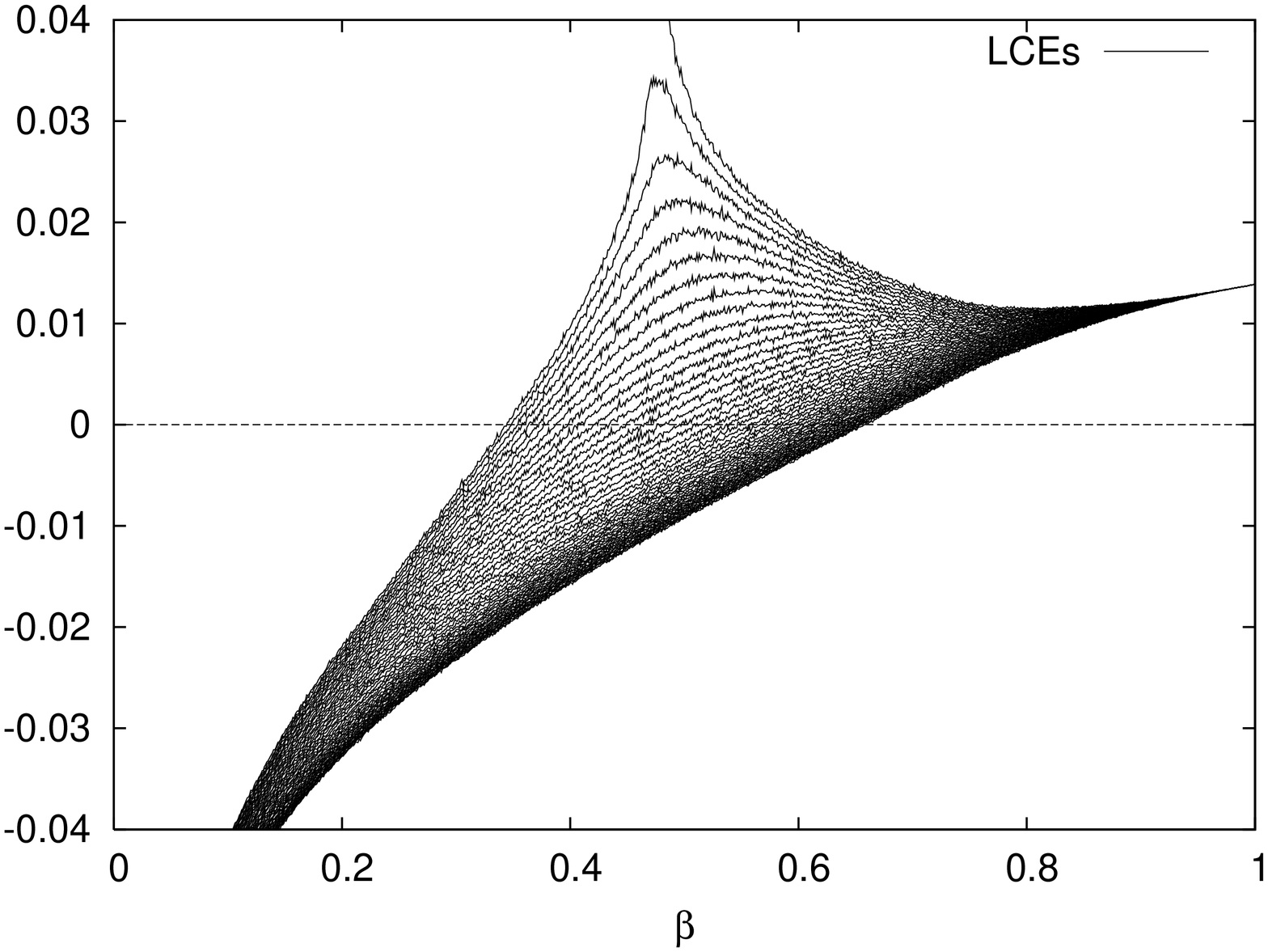 , height=8.0cm, angle=270}
\epsfig{file=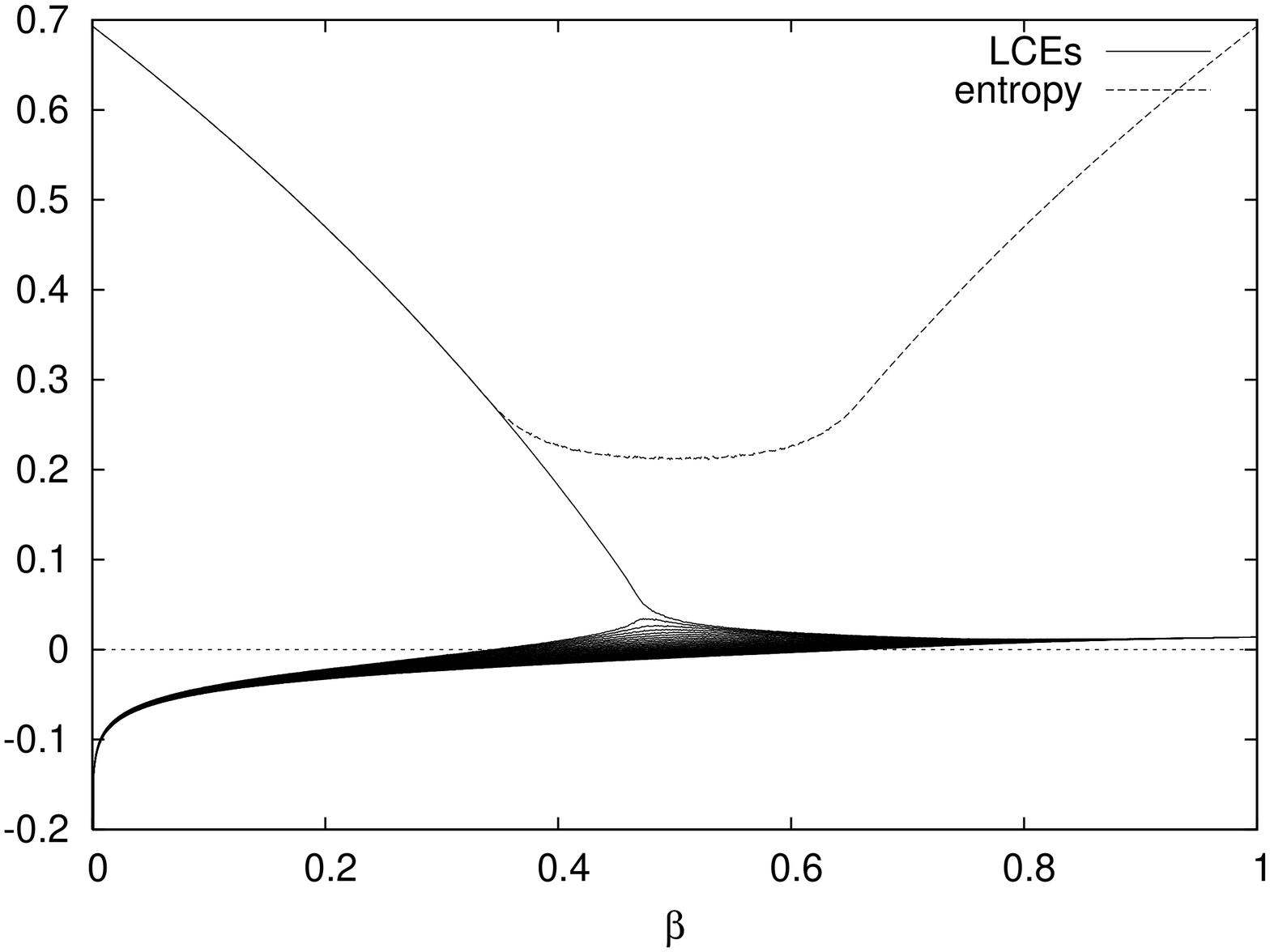, height=8.0cm, angle=270}
\end{center}
\caption{Lyapunov exponents for (\ref{equation:full-delay-sum}) where $f$ is
the standard tent map and $d=\tau+1=50$. The $\beta$ increments are $10^{-5}$, and
for each $\beta$ the LCEs were calculated over $50000$ time-steps. 
The lower subfigure is a zoomed-out version that also shows the entropy plot.}%
\label{fig:tent_2_2_d50}%
\end{figure}

\begin{figure}[tb]
\begin{center}%
\begin{tabular}
[c]{c}%
{\parbox[b]{3.5293in}{\begin{center}
\includegraphics[scale=0.65]{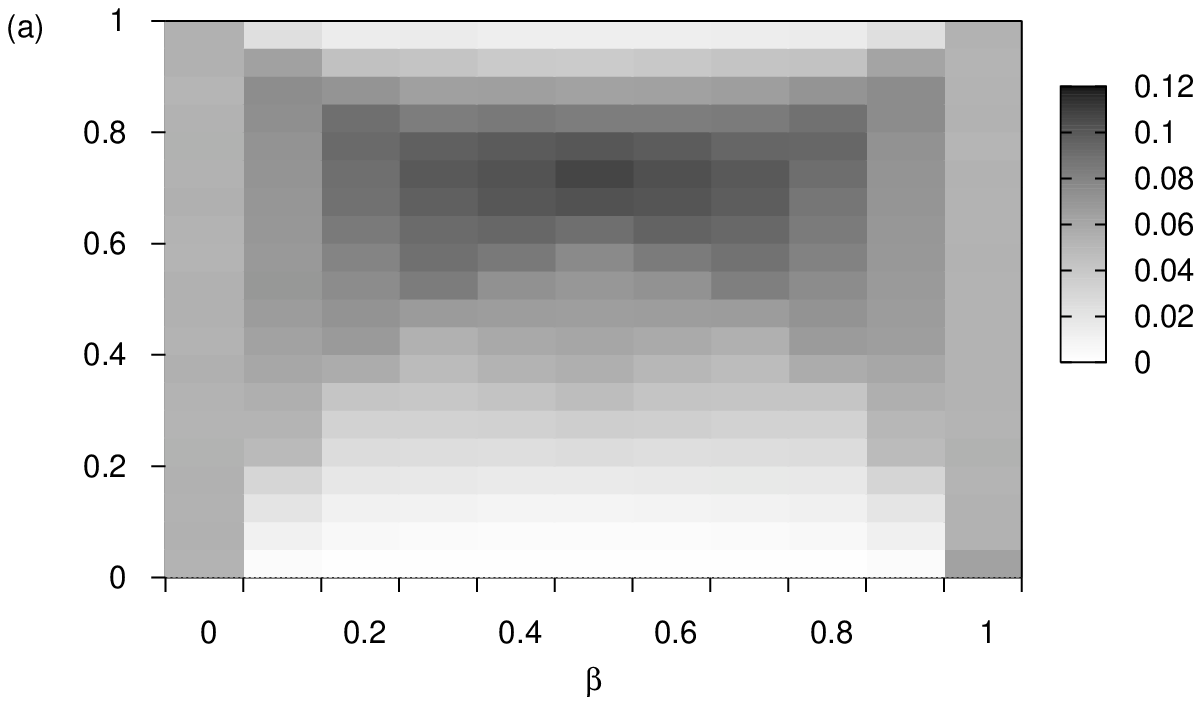}\end{center}}}\\
{\parbox[b]{3.5293in}{\begin{center}
\includegraphics[scale=0.65]{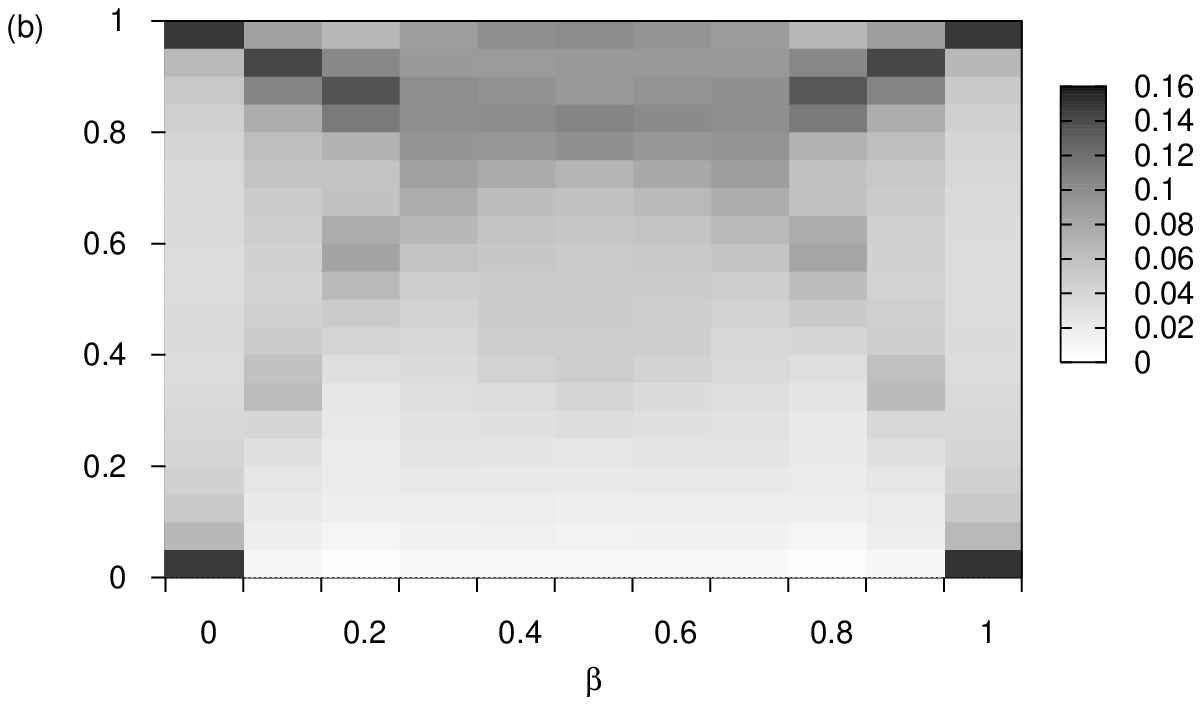}\end{center}}}%
\end{tabular}
\end{center}
\caption{The natural density for (\ref{main}) for (a) the tent and (b) the
logistic maps with delay $\tau=50$. For values of $0\leq\beta\leq1$, the
vertical strips show the relative density of points in the interval $[0,1]$
for the evolution of (\ref{main}), darker shades corresponding to higher
density.}%
\label{fig:density1}%
\end{figure}

Considering Fig.~\ref{fig:tent_2_2_d50}, for ease of description, let
us parse the $\beta$ interval into three dynamical regions with
monotonic ordering as $B_{1}=(0,0.3)$, $B_{2}=(0.3,0.7)$, and
$B_{3}=(0.7,1)$. The first and third regions are transitions to ``pure
states,'' where the dynamics correspond to dynamics of the original
(tent) map with stochastic perturbations, or small perturbations of
the invariant measure. This conclusion is drawn from two observations.
First, given enough time-delays, the diagnostics ($h_{\mu}$ and the LCEs)
in these regions make smooth transitions to their values for the pure
states. Note that in region $B_{3}$ (Fig.~\ref{fig:tent_2_2_d50}), the
LCEs (and thus the entropy and $D_{KY}$) behave in accordance with
Lemma~\ref{lemmma:lle_rescaling}. The
primary difference between regions one and three lies in the different
LCE structure. Nevertheless, considering Fig.~\ref{fig:density1}, the
invariant densities of both region one and three are very similar
(they are seemingly identical). 
Thus, the interpretation of the
dynamics in regions one and three is of original map, $f$, perturbed
by what is essentially (but not technically) noise. One final bit of
support for the claim that regions one and three are dynamically
similar is the observation that for $d>4$, the entropy
(Fig.~\ref{fig:tent_2_2_d50}) and the invariant density
(Fig.~\ref{fig:density1}) are symmetric about $\beta=1/2$.  

\begin{figure}[ptb]
\begin{center}
\epsfig{file=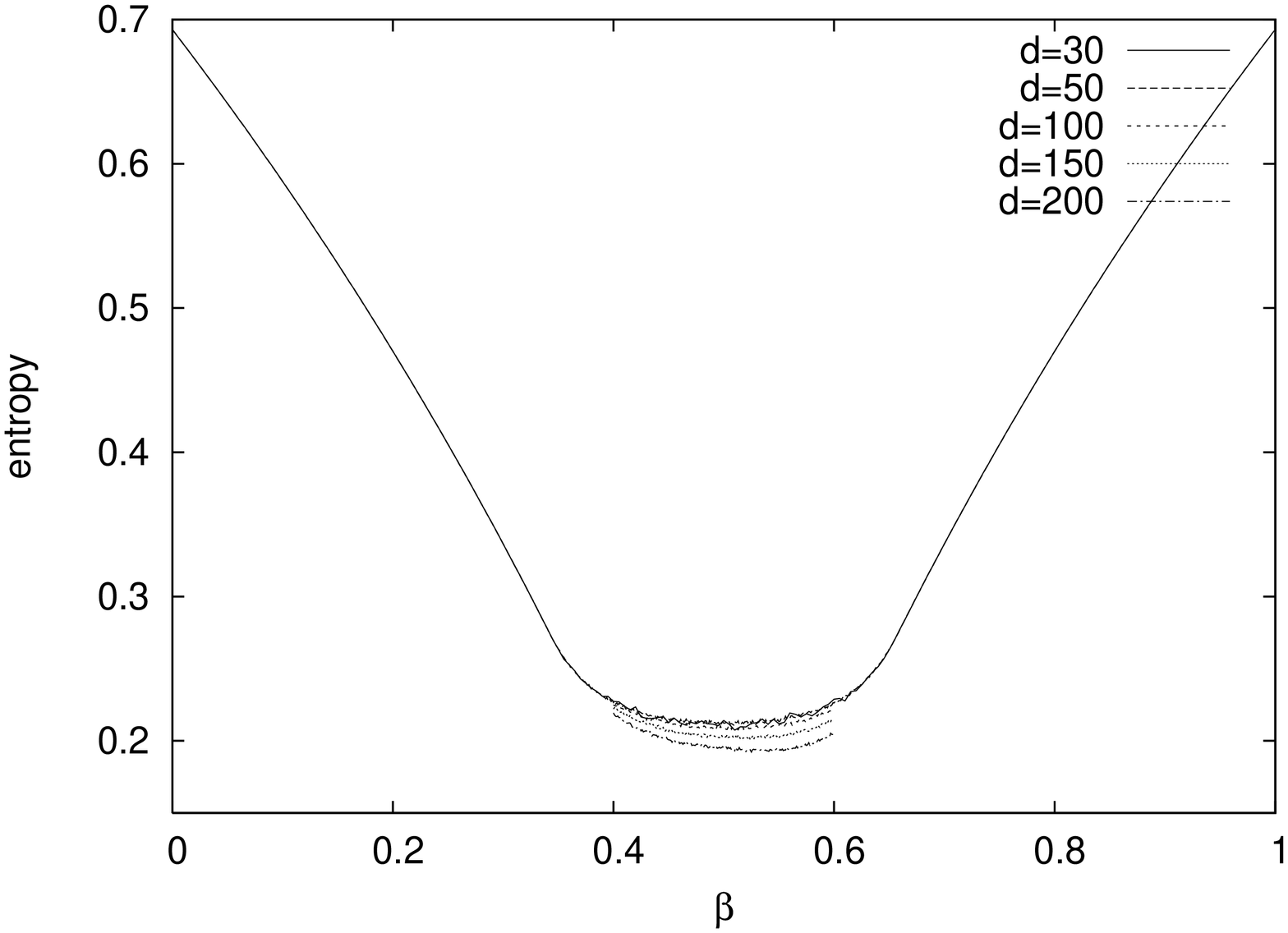, height=8.0cm, angle=270}
\epsfig{file=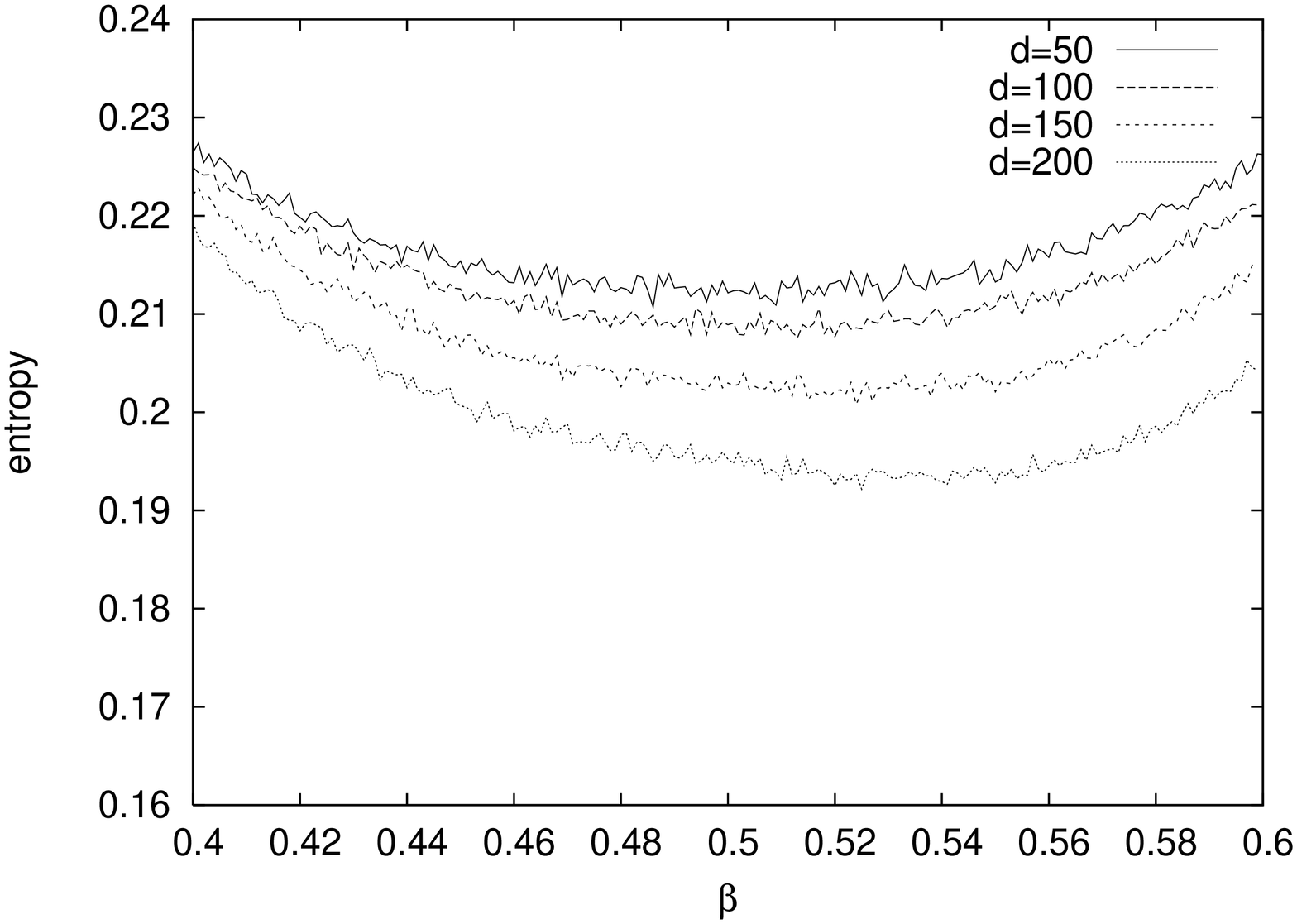, height=8.0cm, angle=270}
\end{center}
\caption{Metric entropy plots of Eq. (\ref{equation:full-delay-sum})
  where $f$ is the standard tent map for dimensions ranging from $30$
  to $200$.  The $\beta$ increments are $10^{-4}$, and for each
  $\beta$ the LCEs were
calculated over $100000$ time-steps.}%
\label{fig:tent_entropy_delay_variation}%
\end{figure}

Region $B_{2}$, we believe, represents a fundamentally different kind
of dynamics from the other regions. It is not a stochastically
perturbed low-dimensional system, nor does it correspond to a
transition to or from the pure states in regions $B_{1}$ or $B_{3}$.
Instead, we claim, based originally on work by Manneville
\cite{mann_lce_scaling} (and a suggestion to the authors by Y.
Kuramoto) that region $B_{2}$ is a representation of a continuous
(LCE) spectrum, akin to a PDE.  This hypothesis is driven by the
qualitative difference in the dynamics that is indirectly witnessed
via two qualitative observations.  Considering the invariant density
as depicted in Fig.~\ref{fig:density1}, it is evident that the
bifurcation that occurs between regions one/three and two leads to a
significantly different invariant density than that of the perturbed
map in regions one and three. This change in the invariant density
suggests that there does exist a fundamental, qualitative difference
between region two and regions one and three.  That this qualitative
change may be independent of dimension above a (soft) threshold can be
seen in the invariance of the entropy.  Considering
Fig.~\ref{fig:tent_entropy_delay_variation}, the entropy is, given $d$
high enough (e.g., $d>20$), largely invariant to increases in
dimension.  In particular, while the dimension is quadrupled, the
change in entropy is less than $10$-percent and well below error
estimates for the given number of iterations used in the numerical
experiments. We assert that the variation in the entropy for
$\beta>1/2$ is a numerical artifact of the errors in the smaller LCEs
or the LCEs near zero.
   
\begin{figure}[ptb]
\begin{center}
\epsfig{file=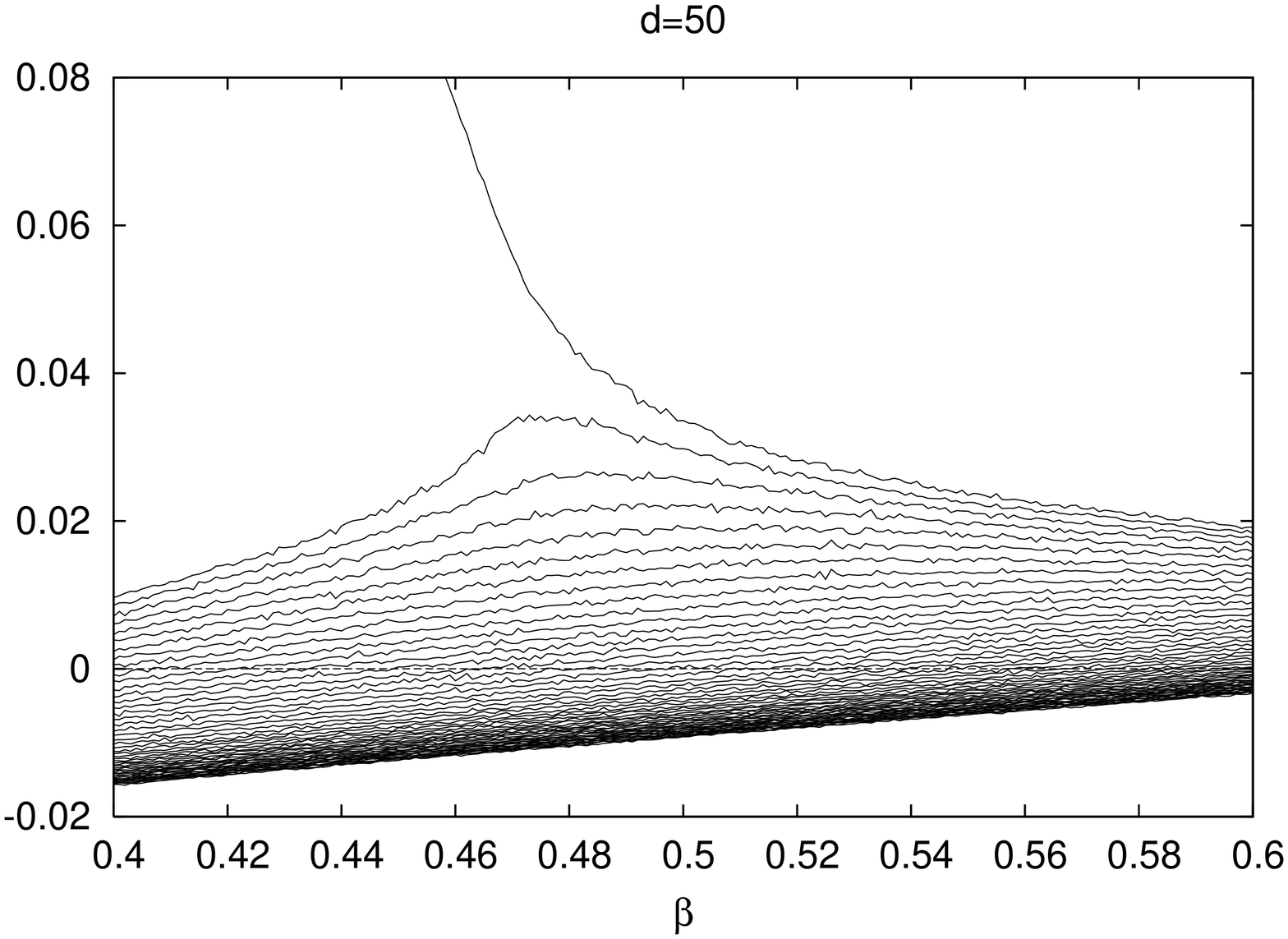, height=7.5cm, angle=270} \epsfig
{file=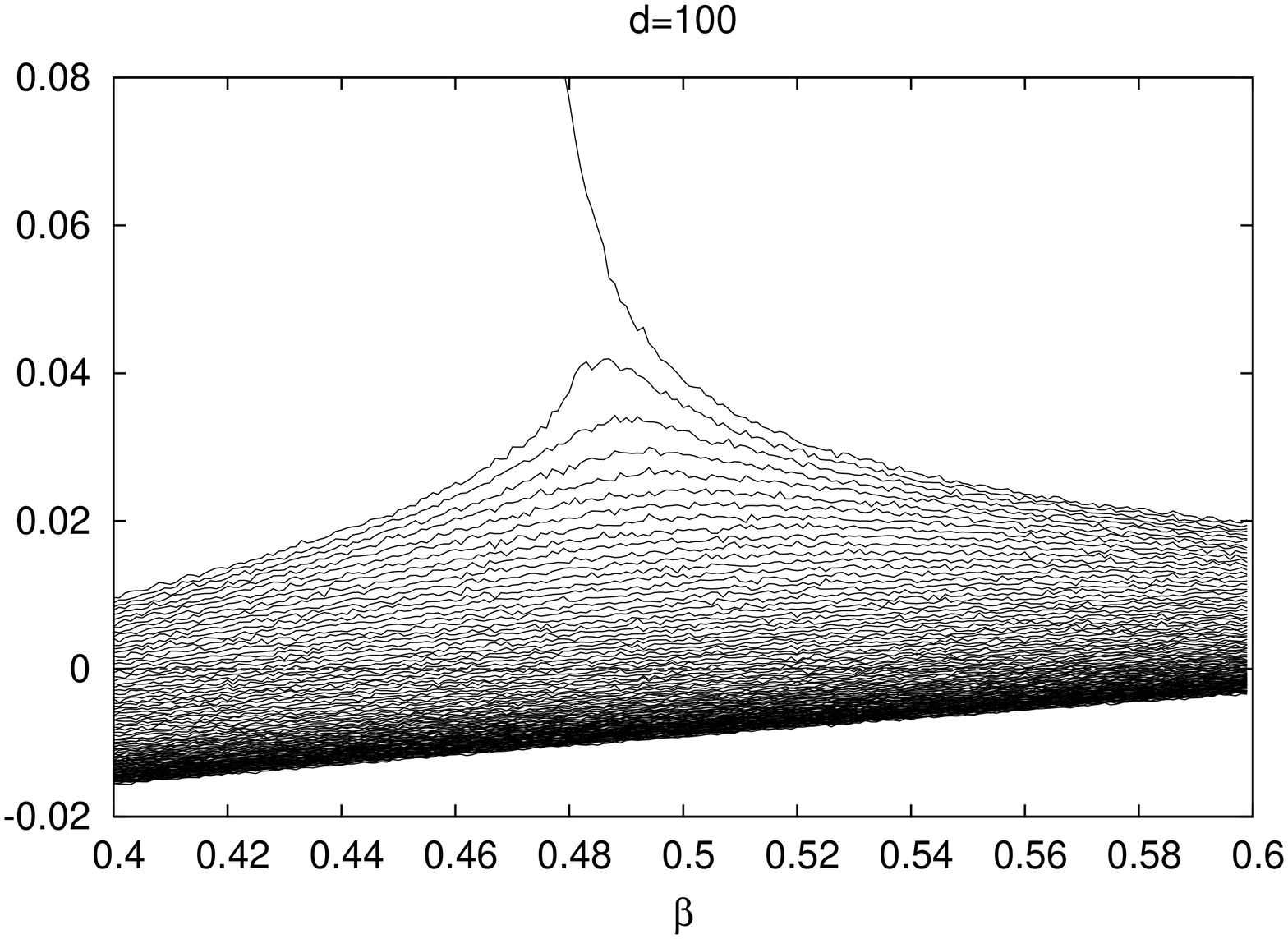, height=7.5cm, angle=270} \epsfig
{file=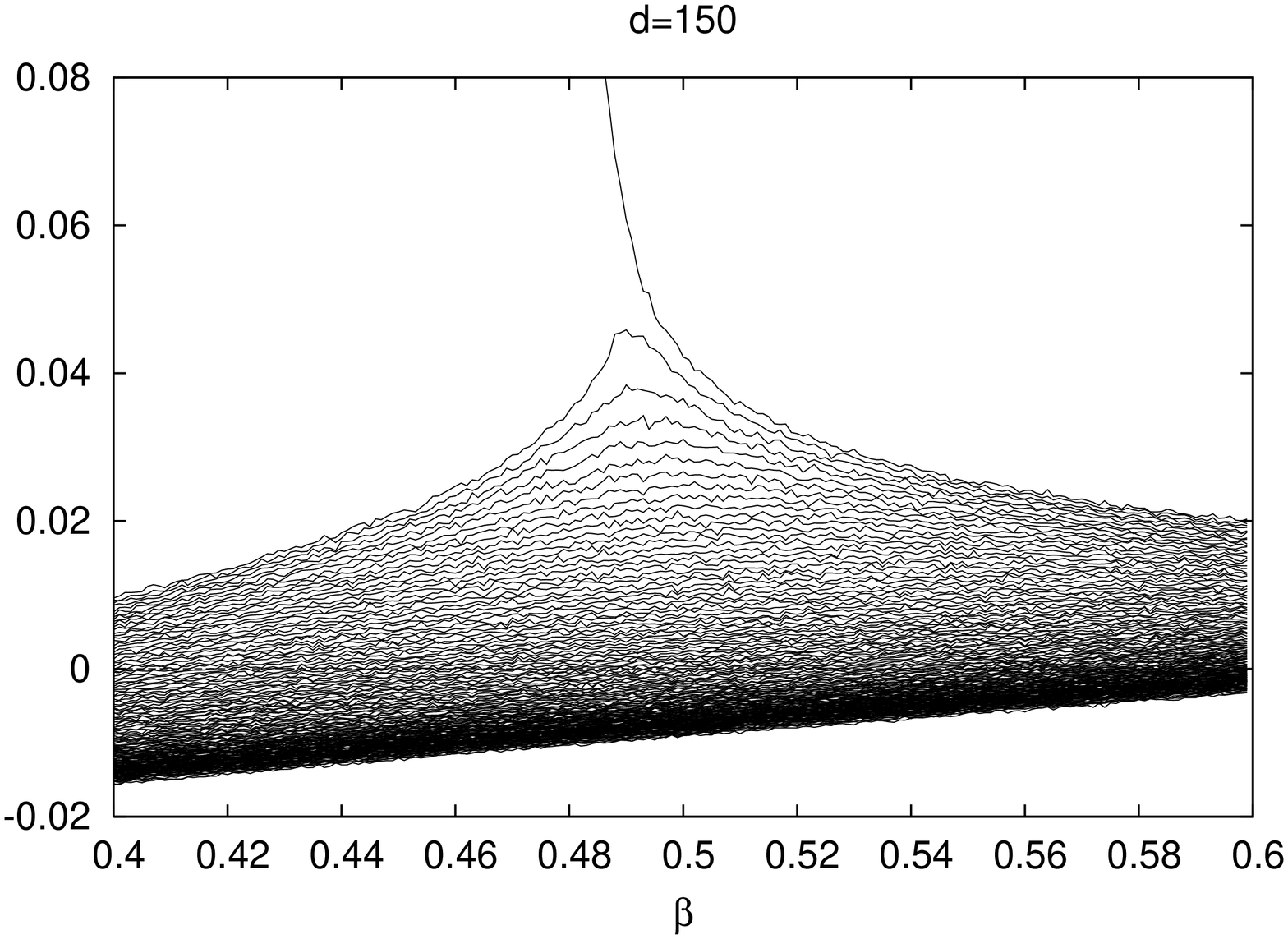, height=7.5cm, angle=270} \epsfig
{file=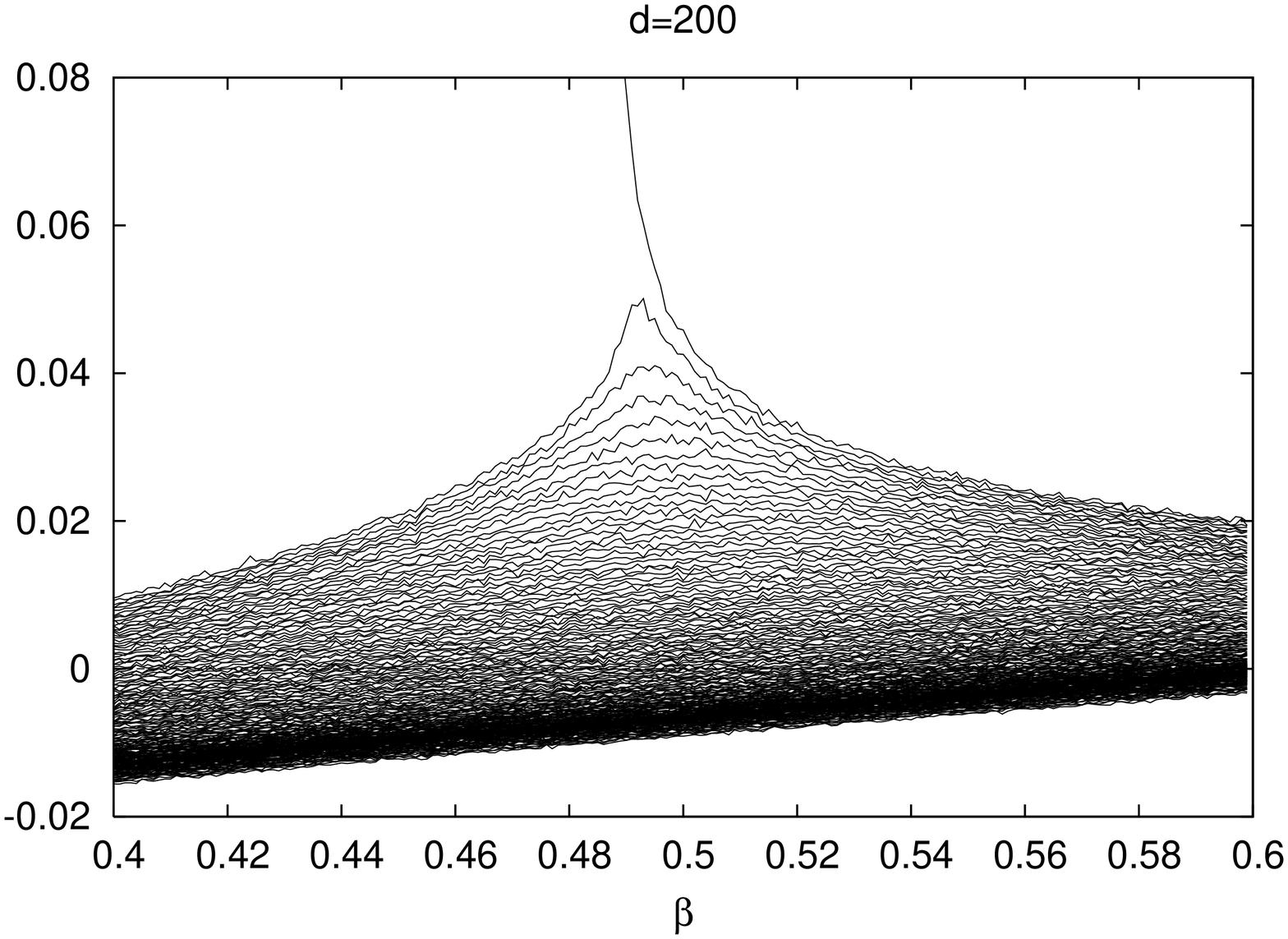, height=7.5cm, angle=270}
\end{center}
\caption{LCE plots of Eq. (\ref{equation:full-delay-sum}) where $f$ is
  the standard tent map restricted to region two ($B_{2}$) for
  dimensions ranging from $50$ to $200$ where the $d>50$ cases have
  been rescaled (by $d/50$) to the $d=50$ time-scale. The $\beta$
  increments are $10^{-4}$, and
for each $\beta$ the LCEs were calculated over $100000$ time-steps.}%
\label{fig:tent_delay_kura}%
\end{figure}

\begin{figure}[ptb]
\begin{center}
\epsfig{file=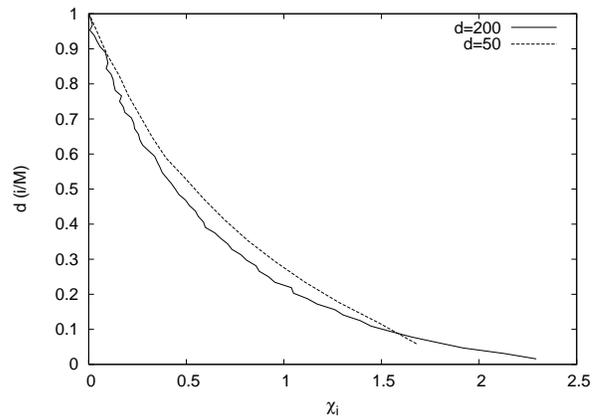, height=8.0cm, angle=270}
\end{center}
\caption{The distribution of LCEs --- $\chi_i$ versus $d
  \frac{i}{M(\beta)}$ --- for Eq. (\ref{equation:full-delay-sum}) where
  $f$ is the standard tent map with dimensions $50$ and $200$ at fixed
  $\beta = 1/2$.}
\label{fig:exponent_distribution_scaling}%
\end{figure}

Quantitative evidence of the existence of a continuous
spectrum type behavior can be gained via a careful consideration of
the LCEs as the dimension is increased.  In particular, considering
the plots in Fig.~\ref{fig:tent_delay_kura} where the LCE spectrum in
region two is displayed for dimensions ranging from $d=50$ to $d=200$,
normalized to $d=50$, the following observation is eminent: upon
adding delays, the LCEs remain distributed in a relatively uniform way
up to a time-rescaling.  In fact, the primary difference in the plots
at different dimensions is that $|\chi_{1}-\chi_{d}|$ decreases with
dimension, and the intermediate LCEs are added in a manner consistent
with their densities at lower $d$ as their numbers are increased.
This statement can be quantified by considering the normalized
distribution of positive LCEs.  To achieve this, we begin by defining
$M(\beta)$ as the number of \emph{positive} LCEs at a given $\beta$.
Next consider the distribution of LCEs, $\mathcal{D}$, via a discrete
plot of $\chi_i$ versus $d (\frac{i(\chi_i)}{M(\beta)})$ (the factor of
$d$ normalizes $\mathcal{D}$ to unity).  As can be seen in 
Fig.~\ref{fig:exponent_distribution_scaling}, there exists a universal
scaling between LCEs that is invariant as $d$ is increased.  Indeed
the least squares fit of
\begin{equation}
d \frac{i(\chi_i)}{M(\beta)} = \alpha e^{\gamma \chi_i}
\end{equation}
for $d=50$ yields $\alpha = 1.08, \gamma = -1.51$ (with a
$\chi^2$-error of $0.978$) whereas for $d=200$, the fit yields
$\alpha = 1.02, \gamma = -1.59$ (with a $\chi^2$-error of $0.992$).
These fits differ by less than five percent over a factor of four in
dimension, and the fitting error decreases considerably with
increasing dimension.  That the LCEs remain relatively uniform (or are
added in a manner consistent with their density for lower-$d$) up to a
time rescaling and increase in dimension, implies that increasing the
number of delays in this region is equivalent to increasing the
resolution in a PDE-like mapping, leading to the conclusion that as
$d\rightarrow\infty$, the LCE spectrum would tend to a continuous
function at fixed $\beta$. If one accepts the proposition that
the``law of large numbers cannot lie,'' this LCE structure is a strong
indication of an invariant (SRB) measure for a continuous-space
system. (An exact qualification of this LCE structure is an object of
future research).

The above reasoning leads us to conjecture that systems with LCE
structure as is seen in $B_{2}$ corresponds to high-entropy,
high-dimensional, equilibrium-like (possibly turbulent-like) systems
that are not easily reduced or approximated by low-dimensional
dynamical systems. Moreover, we believe that the dynamical
characteristics are largely seen as a consequence of exactly the state
mixing that allows the time-series embedding results to work
correctly. It is also interesting that mixing states in some
(non-trivial) circumstances can lead to a highly complicated,
high-dimensional dynamical system. In this case, state mixing leads to
higher-dimensional dynamics than the initial mapping (in this case the
tent map). Finally, the dynamics in region $B_{2}$ can be identified
as having \emph{bifurcation chains} structure, which is defined for an
interval of parameter space such that (i) the number of positive LCEs
increases with increasing $d$ at a given parameter value, (ii) the
Euclidean distance between sequential LCE magnitudes decreases with
increasing $d$ at a given parameter value, (iii) the Euclidean
distance between sequential LCEs remains relatively uniform at a given
parameter value, and (iv) the LCEs cross zero transversally.
Bifurcation chains structure represents a highly irreducible,
high-dimensional dynamic type reminiscent of complex, equilibrium-like
dynamics (such as homogeneous, fully developed fluid turbulence) ---
the bifurcation chains structure is discussed in detail (for a
different system) in Refs. \cite{dynamicsPRL,hypviolation}.

That regions $B_1$ and $B_3$ consist of similar qualitative
dynamics, and that these two regions are separated in $\beta$-space by
region $B_2$ which has qualitatively different dynamics, implies that
the transition between regions $B_1$ and $B_2$, and $B_2$ and $B_3$,
represent a sort of bifurcation, or \emph{phase transition} between
``low-dimensional,'' reducible dynamics and high-dimensional dynamics,
irreducible dynamics.  For the tent map, this phase transition is
quite simple and void of highly complex structure.  As we will see in
the following section, this is likely due to the fact that the tent
map has a nice absolutely continuous invariant measure over all the
parameters we are considering.

Finally, while we refrain from a careful analysis of the dynamics at
$\beta=1/2$, one is tempted to conjecture that this point represents a
bifurcation behavior in parameter space.  It is not only the midpoint
of $B_2$ and thus a turning point of sorts in parameter space, but it
is the point where $D_{KY}$ begins to drop from equality with $d$.
Nevertheless, given that there is no change in the invariant density
of at this point, the bifurcation will have to be characterized in a
novel manner.  It would not be surprising if a homogeneous function,
renormalization style analysis could be performed at this point.

\subsection{Logistic map}

\begin{figure}[ptb]
\begin{center}
\epsfig{file=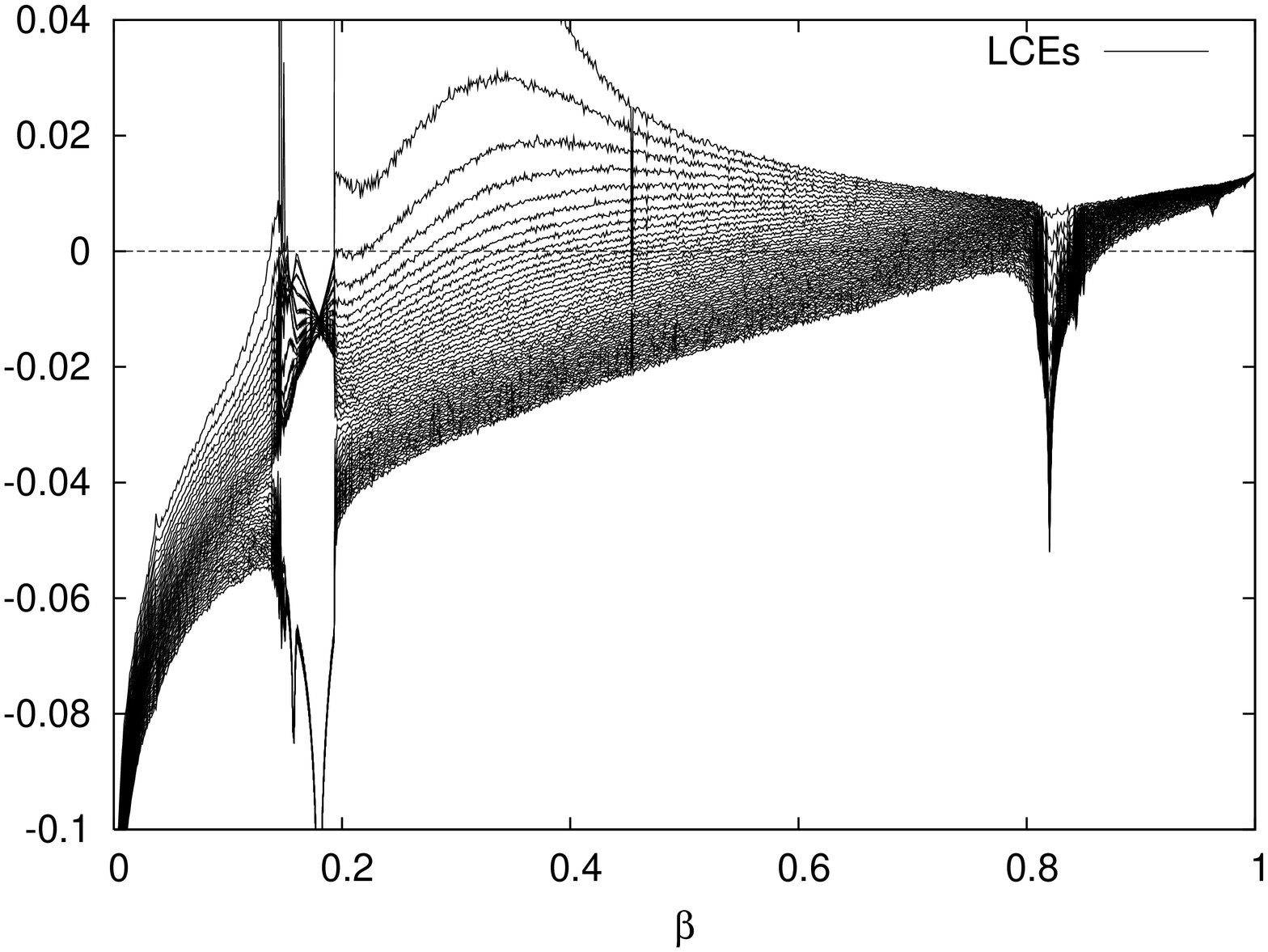, height=7.5cm, angle=270}
\epsfig{file=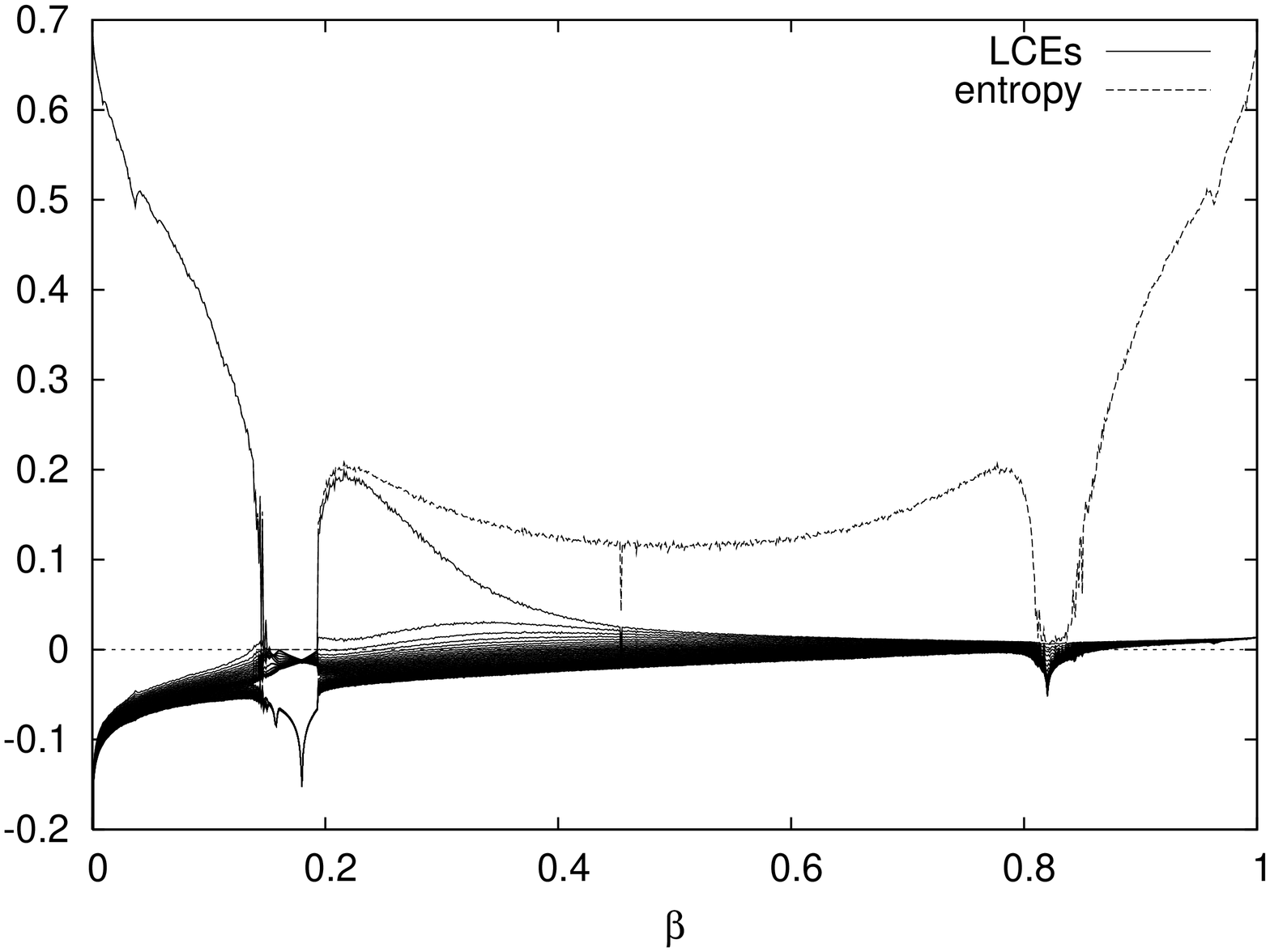, height=7.5cm, angle=270} \epsfig
{file=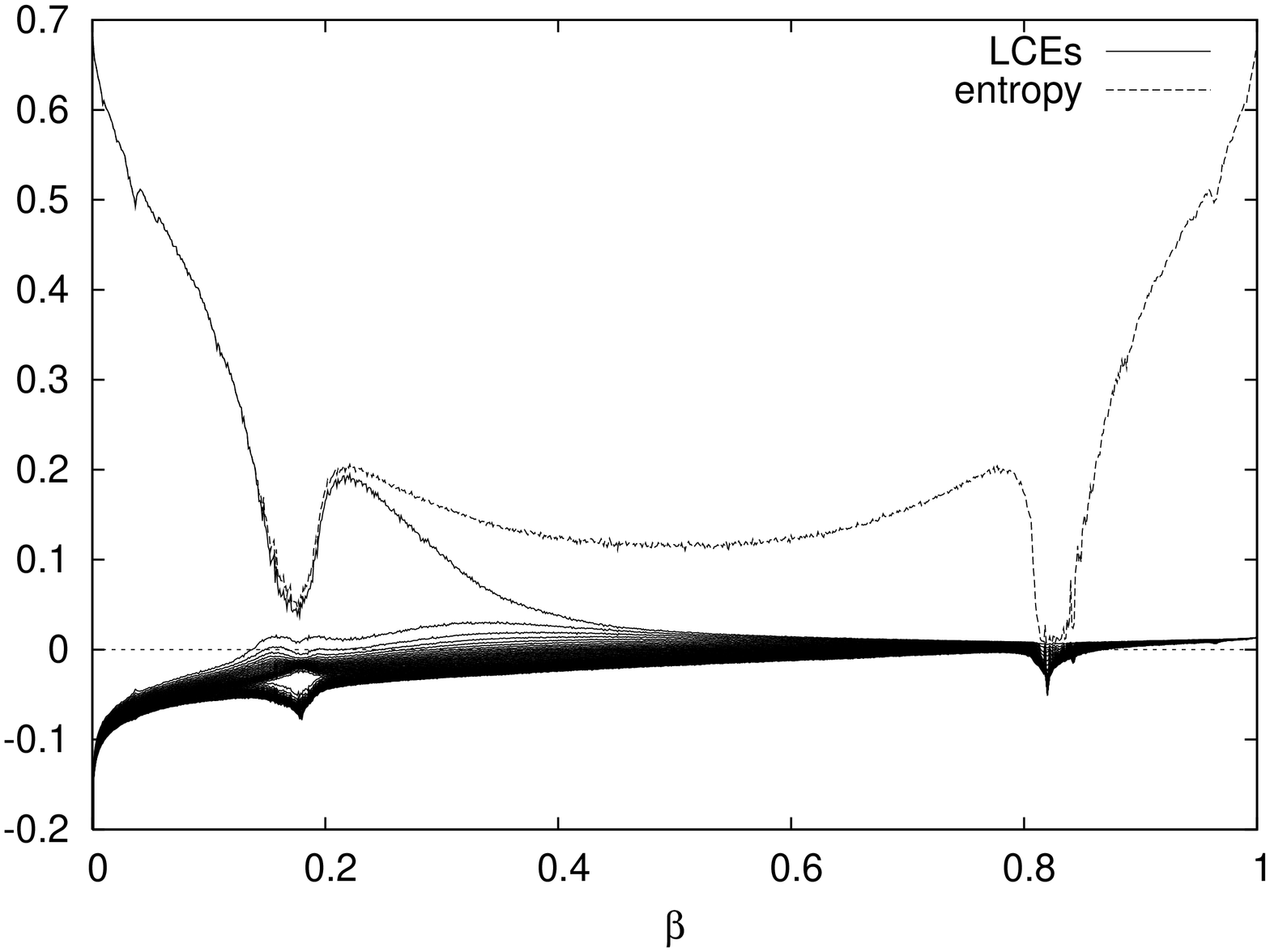, height=7.5cm, angle=270}
\end{center}
\caption{LCE plots of equation (\ref{equation:full-delay-sum}) where $f$ is
the standard logistic map and $d=50$ (top two plots) and $d=51$ (bottom plot).
The $\beta$ increments are $10^{-5}$, and for each $\beta$ the LCEs were
calculated over $50000$ time-steps.}%
\label{fig:logistic_2_2_d50}%
\end{figure}

We now take $f$ to be the standard logistic map given by:
\begin{equation}
f(x_t)=ax_t(1-x_t) \label{equation:logistic}%
\end{equation}
with $a=4$, the parameter setting for which the logistic map is
absolutely continuous \cite{jakobson_abscontin} and is conjugate to
the tent map. Again, for ease of description, let us parse the $\beta$
interval into dynamical regions in monotonic ordering as follows:
$B_{1}=(0,0.15)$, $B_{1\rightarrow 2}^{T}=(0.15,0.2)$,
$B_{2}=(0.2,0.8)$, $B_{2\rightarrow3}^{T}=(0.8,0.85)$, and
$B_{3}=(0.85,1)$. These regions correspond to the case presented in
Figs.~\ref{fig:logistic_2_2_d50} and \ref{fig:density1} where $d$ is
set to $50$. It is worth noting that both Figs.~\ref{fig:density2} and
\ref{fig:logistic_2_2_d50} display a dimension dependence that does
not diminish by simply increasing $d$.

%

Just as was the case for the tent map, the first and third regions are
transitions to ``pure states,'' where the dynamics correspond to dynamics of
the original (logistic) map with stochastic perturbations, or small
perturbations of the invariant measure. Again note that in region $B_{3}$
(Fig.~\ref{fig:logistic_2_2_d50}), the LCEs (and thus the entropy and $D_{KY}%
$) behave as per Lemma~\ref{lemmma:lle_rescaling}. 
There is indeed little difference between
the logistic and tent maps in these regions, which suggests that these
regions will exist and be qualitatively the same for most
\emph{stochastically stable} \cite{ledrappier_young} dynamical systems
if $d$ is large enough. The dynamics seen in regions one and three is
evidence that points to the logistic (for certain parameters) and tent
maps being stochastically stable and satisfying L.-S. Young's
\emph{zero-noise limit} \cite{youngSRB}.

\begin{figure}[ptb]
\begin{center}
\epsfig{file=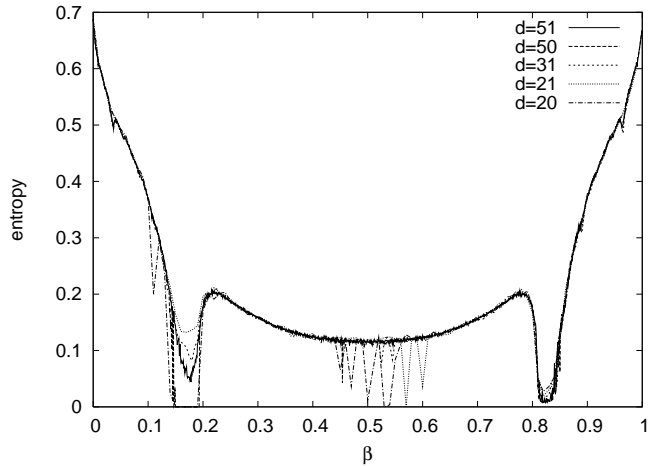, height=9.0cm, angle=270}
\end{center}
\caption{Entropy versus $\beta$ where $f$ is the standard logistic map with a
varying number of delays.}%
\label{fig:logistic_entropy}%
\end{figure}

Region $B_{2}$ is most easily seen by considering either the invariant
density in Fig.~\ref{fig:density1} (where the invariant density
changes little between regions $B_{1\rightarrow2}^{T}$ and
$B_{2\rightarrow3}^{T}$ yet is qualitatively distinct from regions
$B_{1}$ and $B_{3})$, or the LCE spectrum in
Fig.~\ref{fig:logistic_2_2_d50} where the bifurcation chains structure
appears. Indeed, region $B_{2}$ is roughly the same for the logistic
and tent maps, and we impart a similar interpretation of the dynamics.
Nevertheless, there are important differences. For lower dimensions,
the logistic map does display small periodic windows in region two, as
can be seen in Fig.~\ref{fig:logistic_entropy}.  The state mixing
combined with added dimensions appears to have the effect of
destroying the stable periodic orbits if $d$ is large enough ---
periodic orbits are observed for $d\leq30$, whereas for $d\geq50$, if they
exist, they are below the $\beta$ resolution of $10^{-5}$.  It is
possible that the difference between the logistic and tent maps is a
combination of the fact that the logistic map does not have persistent
dynamics relative to parameter perturbations (i.e., the existence of
dense, stable periodic orbits for $a\in\lbrack0,4]$) contrasted with
the relative dynamical persistence of piecewise smooth maps
\cite{yorkerobustchaos} such as the tent map.

The \emph{$f$-dependence} appears profoundly in the phase transition
regions, $B_{1\rightarrow2}^{T}$ and $B_{2\rightarrow3}^{T}$.  The
structure of the transitional regions between the low-dimensional
``pure states'' and the high-dimensional dynamics of region $B_{2}$
are particular to the logistic map.  In particular, both regions
correspond to an effective value of the parameter
$a_{\mathrm{eff}}\in(3.23,3.45)$ (where $a_{\mathrm{eff}}=\max\{\beta
a,(1-\beta)a\}$), which corresponds to the region between the
bifurcation from a fixed point to period two, but before the
bifurcation from period two to period four, of the logistic map. The
boundaries of these regions are roughly independent of magnitude of
the dimension (as can be seen by considering the entropy versus
dimension shown in Fig.~\ref{fig:logistic_entropy}), but these regions
\emph{do} have a dimension parity dependence. Assuming $d>4$, region
$B_{2\rightarrow3}^{T}$ is never a periodic window independent of the
dimension parity. In contrast, region $B_{1\rightarrow2}^{T}$ is not a
periodic window when $d$ is odd but is \emph{always} a periodic window
when $d$ is even. Moreover, while the width of regions
$B_{1\rightarrow2}^{T}$ and $B_{2\rightarrow3}^{T}$ are roughly
equivalent and symmetric about $\beta_{\tau}=1/2$, they have different
shapes and structures.  This implies that if $\nu$ is a random
variable with the invariant measure of $f(x_{t-\tau})$, both
$f(x_{t})+\nu\neq f(x_{t}+\nu)$ and $f(x_{t})+\nu\neq
f(x_{t-\tau})+\nu$, when $f$ is the logistic map. (In contrast to the
logistic map, it appears that the time-ordering does not matter for
the tent map.)

\section{Summary}

Putting all the pieces together, for the time-delay systems 
(\ref{main}) and (\ref{PD}), if the number of
dimensions, $d=\tau + 1$, is large enough, entropy remains roughly invariant
to increases in $d$ while the LLE, the LCEs, and $D_{KY}$ do not.
While the LLE and LCEs can still yield insight into the global
structure of the attractor, many dimension calculations such as the
Kaplan-Yorke dimension may yield deceiving results. We conjecture this
is in general true for systems of the form (\ref{equation:full-delay-sum}),
 largely because dimension calculations have an implicit
dimension- and thus coordinate dependence.  Because of these issues, it
is likely that diagnostics such as the metric entropy or the
statistical complexity \cite{Crut88a}, which are truly independent of
coordinates, will be more useful for showing equivalence and
difference in time-delay dynamical systems. Beyond the analysis of the
diagnostics used to describe and investigate time-delay systems, we
also demonstrated that both the time unscaled map with elements of the
distant past and the time rescaled map with the elements of the
current state produce roughly similar dynamics reminiscent of the
$1$-d map plus noise. But, as the distant past and current states are
mixed in more equal parts, the mixing of states only separated with
time-delays can give rise to high-dimensional, irreducible, chaotic
dynamics that we claim can approximate a PDE-like system if the mixing
is via nearly equal contributions of states, and there exist enough
degrees of freedom manifested as time-delays. Thus, we demonstrate two
distinct classes of dynamics: one where the dynamics represent an
infinite-dimensional system; and one where the dynamics represent a
finite-dimensional system, with a phase transition (bifurcation)
between the two dynamical classes, all in the simple context of mixing
only two states of a single mapping. Moreover, this PDE-like dynamics
produces a great deal of dynamic stability even for mappings that have
a lot of periodicity without delays; thus the non-trivial state mixing
can produce relatively stable chaotic dynamics over a sizable interval
in parameter space.  We hypothesize that this dynamic stability
(persistence of chaos) occurs when the delay times allow for enough
de-correlation between the active (non-zero) terms of
Eq.~(\ref{equation:full-delay-sum}) to mix states in a non-linear, but
non-random-like manner. Nevertheless, the dynamics are dependent on
the original maps that compose the time-delay. Finally, in
\cite{dynamicsPRL} and \cite{hypviolation}, an example of bifurcation
chains structure was presented that, relative to a measure on a
function space, was persistent to parameter perturbations. Moreover,
in these examples, in the presence of the bifurcation chains
structure, the probability of periodic windows decreased as dimension
increased. Here we observe similar results, but note that the
bifurcation chains alone do not imply stability or lack of periodic
windows as can be seen via the middle plot of Fig.~\ref{fig:logistic_2_2_d50}.

D. J. Albers wishes to thank J. Dias, J. Jost, Y. Kuramoto, Y. Sato, C. R. Shalizi, J.
C. Sprott, and U. Steinmetz for helpful discussions.%

\bibliographystyle{plain}
\bibliography{ekydtr}

\end{document}